\documentclass{aastex63}

\usepackage{amsmath}
\usepackage{graphicx}
\usepackage{natbib}
\usepackage{lineno}

\shorttitle{High Contrast, High Resolution Speckle Imaging}
\shortauthors{Howell et al.}

\begin{document}

\title{High Contrast, High Angular Resolution Optical Speckle Imaging: \\
Uncovering Hidden Stellar Companions}


\correspondingauthor{Steve B. Howell}
\email{steve.b.howell@nasa.gov}

\author[0000-0002-2532-2853]{Steve~B.~Howell}
\affiliation{NASA Ames Research Center, 
Moffett Field, CA 94035 USA}

\author[0000-0002-3311-4085]{Arturo O.~Martinez}
\affiliation{NASA Ames Research Center, Moffett Field, CA 94035 USA}

\author[0000-0000-0000-0000]{Douglas A.~Hope}
\affiliation{Georgia Tech Research Institute, 925 Dalney St., Atlanta, GA, 30318, USA}
\affiliation{Georgia State University, 25 Park Place, Atlanta, GA 30303,USA}

\author[0000-0002-5741-3047]{David R.~Ciardi}
\affiliation{NASA Exoplanet Science Institute, Caltech/IPAC, Mail Code 100-22, 
1200 E. California Blvd., Pasadena, CA 91125, USA}

\author[0000-0002-9580-5615]{Stuart M.~Jefferies}
\affiliation{Georgia State University, 25 Park Place, Atlanta, GA 30303,USA}

\author[0000-0000-0000-0000]{Fabien R.~Baron}
\affiliation{Georgia State University, 25 Park Place, Atlanta, GA 30303,USA}

\author[0000-0003-2527-1598]
{Michael B.~Lund}
\affiliation{NASA Exoplanet Science Institute, Caltech/IPAC, Mail Code 100-22, 
1200 E. California Blvd., Pasadena, CA 91125, USA}

\begin{abstract}
We explore the possibility of detecting very faint, very close-in stellar companions using large aperture ground-based telescopes and the technique of optical speckle imaging.
We examine the state of high angular resolution speckle imaging and contrast levels being achieved using current speckle cameras on the Gemini 8-m telescope. We then explore the use of the modern image reconstruction technique - Multi-Frame Blind Deconvolution (MFBD) - applied to speckle imaging from the Gemini 8-m telescope. We show that MFBD allows us to measure the flux ratio of the imaged stars to high accuracy and the reconstructed images yield higher precision astrometry. Both of these advances provide a large refinement in the derived astrophysical parameters compared with current Fourier techniques. MFBD image reconstructions reach contrast levels of $\sim$5$\times$10$^{-3}$, near the diffraction limit, to $\sim$10$^{-4}$ about 1.0 arcsec away. 
At these deep contrast levels
with angular limits starting near the 8-m diffraction limit ($\sim$20 mas), most stellar companions to a solar-like stars  can be imaged in the optical to near-IR bandpass (320-1000 nm).
\\
{\it{``To Xanadu we go...''} -- adapted from S. T. Coleridge}
\end{abstract}



 \section{Introduction \label{sec:Intro}}

Speckle interferometry, as a technique for high-resolution optical imaging, began in 1970 with the initial work of \citet{labeyrie:1970}. Using a coherent light source to analyze short exposure photographic images of the ``speckle'' patterns produced by starlight passing through the Earth's atmosphere, Labeyrie was able to show that such observations could remove the effects of seeing induced fluctuations due to distortions in the wavefront and reach the diffraction limit of the telescope. This early work is highlighted in  \citet{gezari:1972} and \citet{labeyrie:1974} who present observational results obtained at Mt.~Palomar revealing resolved stellar disks and close binaries.

Numerous speckle interferometric studies have been performed since the early 1970's, using the better detectors of the day such as photomultiplier tubes, video tubes, and reticons \citep[e.g.,][]{bonneau:1980,mcalister:1987,weigelt:1985,horch:1992,balega:1993}. Obtaining higher S/N digital values for the images, compared with photographic results, allowed for successful speckle image reconstructions using Fourier analysis and interferometric image reconstruction techniques. Bright binary stars have long been the primary speckle research activity over the years, with repeated imagery producing precise stellar orbits \citep[e.g.][]{mcalister:1989}.
Many dozens of papers using speckle interferometry were written in the 1980's and 1990's and produced image reconstructions based on the original, decades old interferometric Fourier-type solutions originally suggested by \citet{labeyrie:1970}. Fourier-based image reconstructions are often enhanced today with the bispectrum technique developed by \citet{Weigelt1977:OptCo..21...55W} and \citet{lohmann:1983}.

Significant advances in astronomical detectors, in particular the higher quantum efficient charge-coupled devices (CCDs), provided a leap forward in this field, allowing for photon intensification, digital outputs, and higher S/N observations to be obtained \citep[e.g.,][]{mcalister:1989,mason:1997,hartkopf:2000,horch:2000}. In recent years, the introduction of the electron multiplier CCD (EMCCD) as a detector \citep{tokovinin:2008,horch:2011b} has been a game changer. With their ability for ultra-fast readout, having essentially zero read noise, near perfect high quantum effeciency, optical flatness, and ease of use, EMCCDs have revolutionized speckle imaging. Fainter astronomical targets can now be observed \citep{howell:2021c}, the overall data quality and signal-to-noise (S/N) of the observations are greater, and the final reconstructed images have more fidelity \citep[e.g.,][]{howell:2022}.

This renaissance in speckle imaging has led to the development of new dedicated speckle instruments \citep[e.g.,][]{howell:2021a,clark:2020,pedichini:2016}, a broadening of the observational science goals, and the placement of such instruments on the worlds largest telescopes \citep{horch:2012,wooden:2018,scott:2021,howell:2021c}. Speckle work is no longer limited to just bright star astrometry but has expanded into many areas of point source and non-point source imagery providing astrophysical information \citep[e.g.,][]{Ricardo2020RNAAS...4..143S,scott:2021,SHARA2022MNRAS.509.2897S}. IR/AO systems typically reach $\sim$0.07$\arcsec$ resolution at K-band and often use some type of occulting disk or coronagraph to block out bright star light in order to achieve high contrasts and search for close, faint companions. Speckle imaging does not need any starlight suppression and routinely reaches an inner working angle (IWA) near the diffraction limit of the telescope \citep[e.g., 0.02$\arcsec$ at 600~nm for an 8-m;][]{lester:2021}.

A second major advance of the past few years is the development of new software techniques for speckle image reconstruction. The analysis of a series of images with distorted wavefronts (i.e., speckle images) can be approached as a deconvolution problem. \citet{jefferies:2011} and \citet{hope:2022} have pioneered wavefront deconvolution techniques to better estimate the atmospheric distortions and produce high quality image reconstructions.  MFBD is a post-processing technique using software algorithms to approximate what deformable mirrors and feedback loops do in IR adaptive optics (AO) systems, however it is far less costly to implement. MFBD is especially useful in the optical bandpass where observations will yield higher angular resolutions. 

This paper examines current speckle imaging using EMCCD detectors on 8-m class telescopes, the analysis of these data using our implementation of Fourier interferometric techniques \citep{howell:2011,horch:2015}, and the MFBD algorithms for image reconstruction. We will show that by using MFBD techniques, robust milestones in achieved contrast levels and the accuracy of the flux ratios are reached using current instrumentation on the Gemini 8-m telescope. Contrasts of $\ge$10$^{-3}$ (8 magnitudes) at the diffraction limit are achieved, flux ratio accuracy of
$\pm 0.03$ magnitudes are obtained, and truly diffraction limited images can be produced.  All the Gemini 8-m `Alopeke speckle data used in this paper is publicly available at the NASA Exoplanet Archive\footnote{https://exoplanetarchive.ipac.caltech.edu/}.

\section{Synopsis of Speckle Interferometry} \label{sec:spec}

Speckle imaging performed in the optical band-pass (0.35-1.0 microns) provides the highest angular resolution available today on any single telescope, delivering 2-4 times better angular resolution than IR/AO observations at K-band. 
Speckle imaging with large (4-8 m) ground-based telescopes routinely observe stars from very bright to as faint as R=16+ and obtain contrasts near 5 magnitudes at 0.2 arcsec and $\sim$8 magnitudes near 1.0 arcsec \citep{howell:2016, howell:2021b}.  Today, for observations performed on 8-m class telescopes under superb observing conditions we find that the angular resolution meets or 
can exceed the diffraction limit \citep{horch:2012} and achieved contrasts can be large \citep{hope:2019,howell:2022}.

Speckle imaging of a target requires many thousands of short  exposures (10-60 ms, depending on the wavelength dependent atmospheric coherence time, that is the observing night conditions) to be obtained and processed. A large number of images are required in order to build up sufficient S/N especially at contrasts greater than 10$^{-3}$ at very close angular separations. While this number of images seems daunting in terms of observing time, at tens of milliseconds per image, typical speckle observations last only a few to a few tens of minutes per target \citep{hope:2019,howell:2021b}. A recent summary of speckle imagers in astronomy was presented in \citet{scott:2021}.

\subsection{Speckle Data Reduction and Image Reconstruction}

Nearly all currently used speckle image reconstruction software packages are based on Fourier speckle interferometric (SI) methods \citep[see][]{labeyrie:1970,lohmann:1983,horch:2015}. The Fourier results presented in this paper are based on our implementation of Fourier methods as described in \cite{howell:2011} and \cite{horch:2012}. Fourier speckle deconvolution techniques are computationally efficient but with the downside that they need additional observations or information about the point spread function of the telescope convolved with the instrument in order to ``calibrate" the restored object’s spectral amplitudes. That is, for point source observations, an additional single (``PSF standard") star is often observed near in position and time to the target star. Sometimes this reference point source does not provide a very good match to the target point-spread function due to changing atmospheric conditions, color differences between the two stars, or the reference star is itself an unknown close binary. 

\begin{figure*}
\centering
\includegraphics[width=0.45\textwidth]{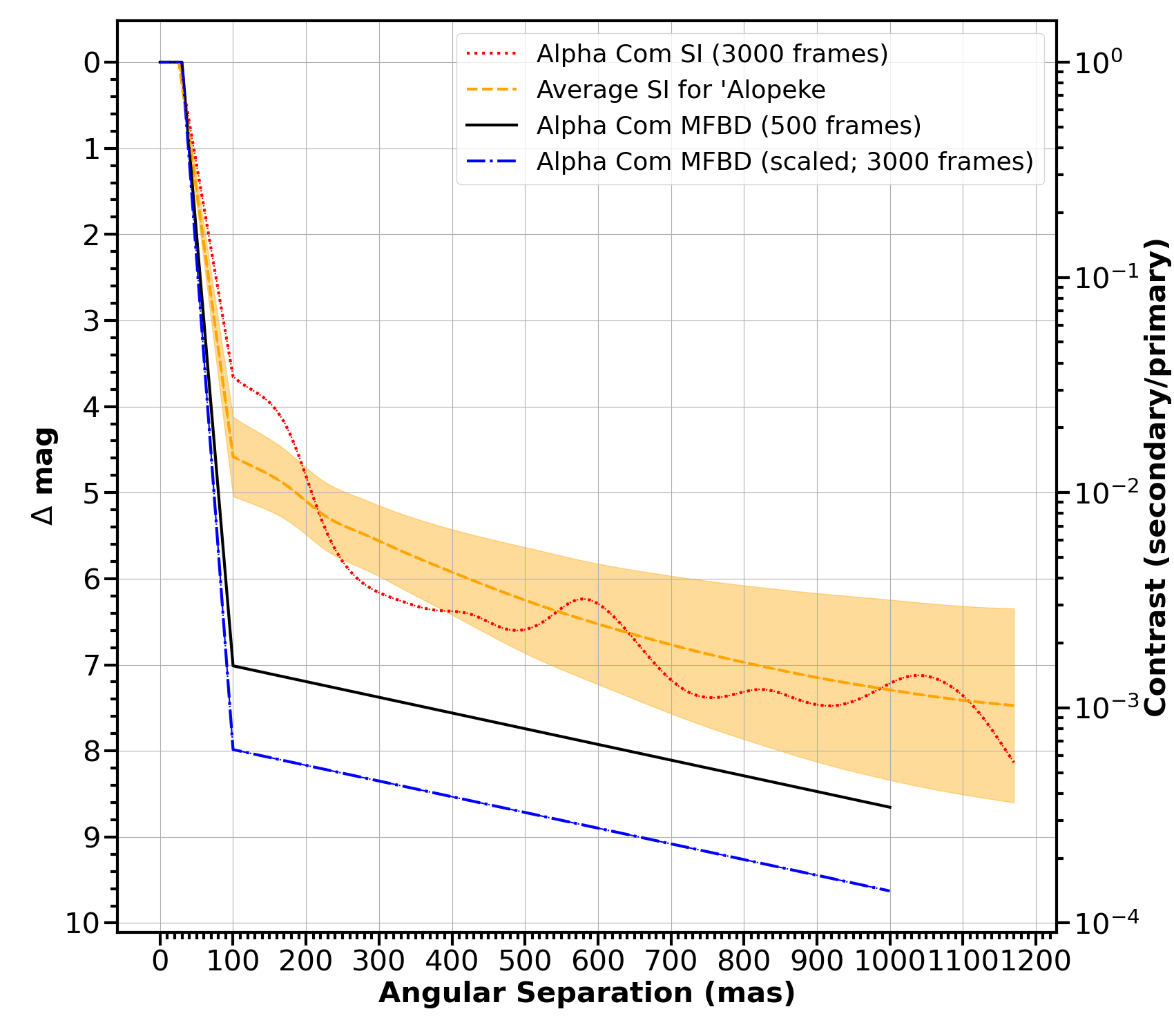}
\caption{
We present contrast curves for the binary system, $\alpha$ Com as well as an average contrast curve for `Alopeke, obtained using publicly available `Alopeke speckle images archived at the NASA Exoplanet Archive. The contrast curves are based on our standard speckle interferometric (SI) pipeline (see \S2.1) and our new results using MFBD as described herein. The red line shows the 5-$\sigma$ contrast curves from SI using a total of 3 minutes of data for $\alpha$ Com (3000, 60~ms frames). The orange dashed line shows the average SI contrast curve with a 1-$\sigma$ deviation (shaded region) for `Alopeke observations from 2019 to mid-2020 \citep[based on data from][]{scott:2021}. The black curve shows the MFBD algorithm results for $\alpha$ Com speckle observations using the best 500 frames of data (i.e., 30 seconds of data). The blue curve shows a S/N scaled contrast expected from the MFBD algorithm for 3000 best frames. Note the larger contrast achieved with MFBD especially near the inner working angle (i.e., the 8-m diffraction limit). All observations were made at 832~nm. See \S3.2.2.
\label{fig:contrastcurve_all}}
\centering
\end{figure*}

\begin{figure}
\centering
\includegraphics[width=0.40\textwidth]{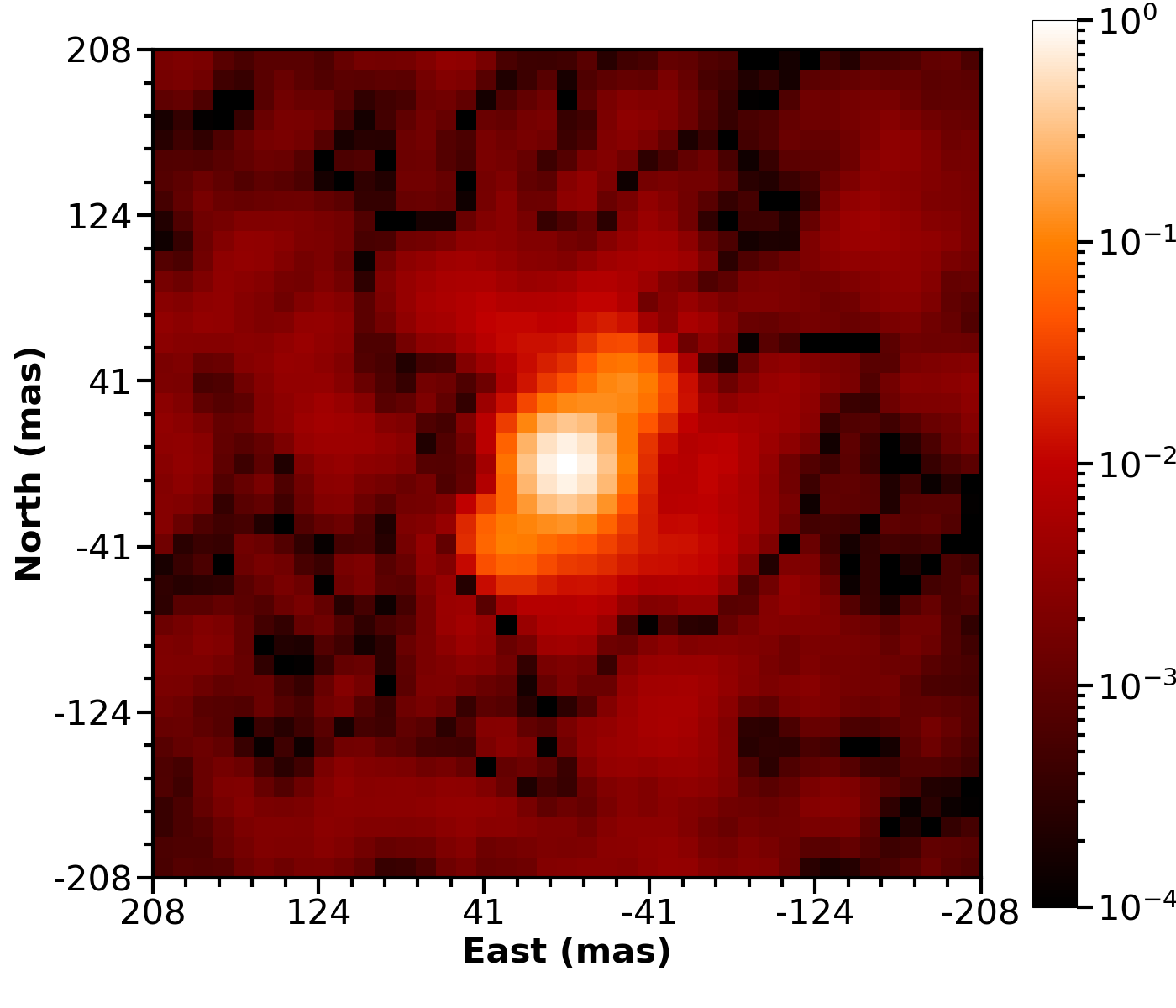}
\includegraphics[width=0.40\textwidth]{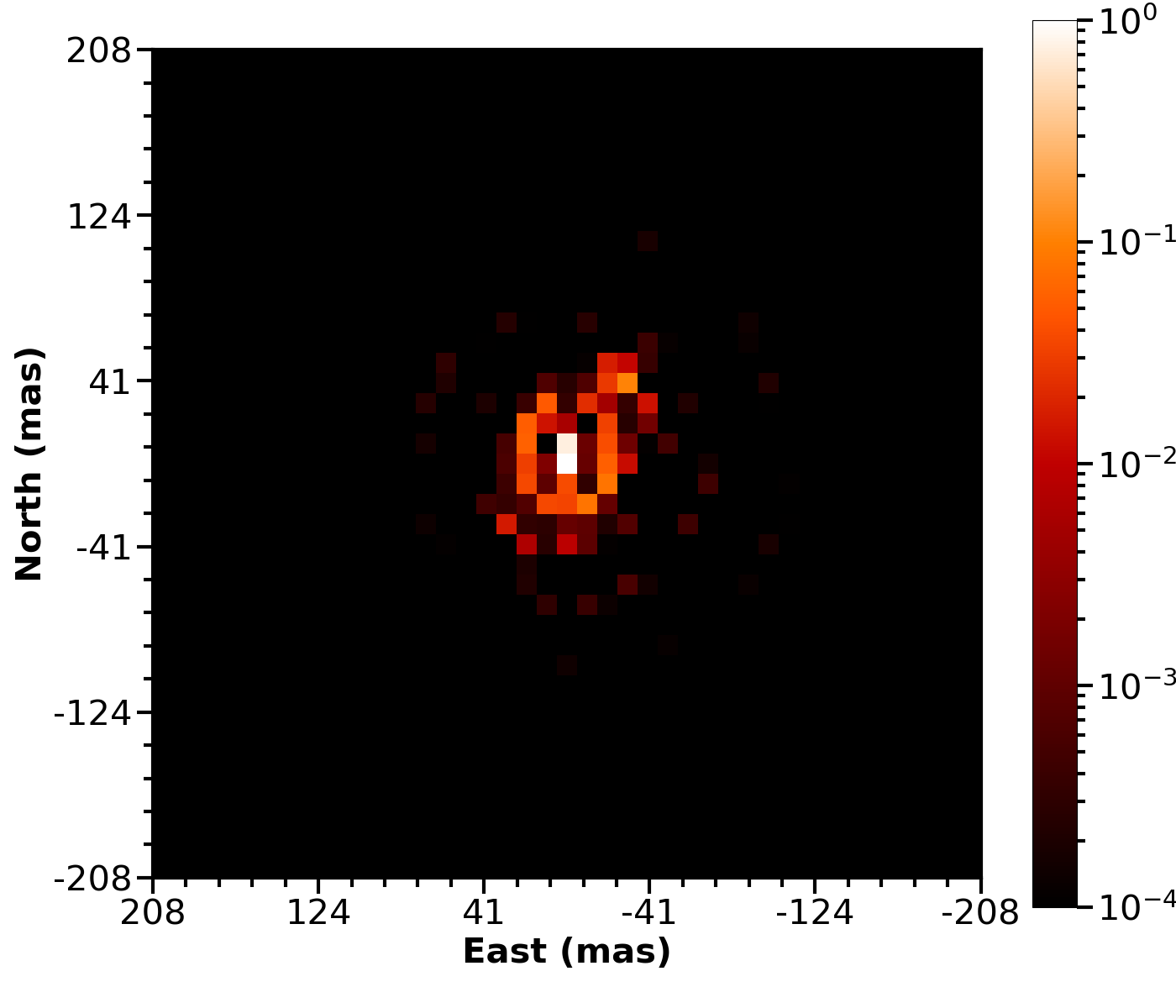}
\includegraphics[width=0.40\textwidth]{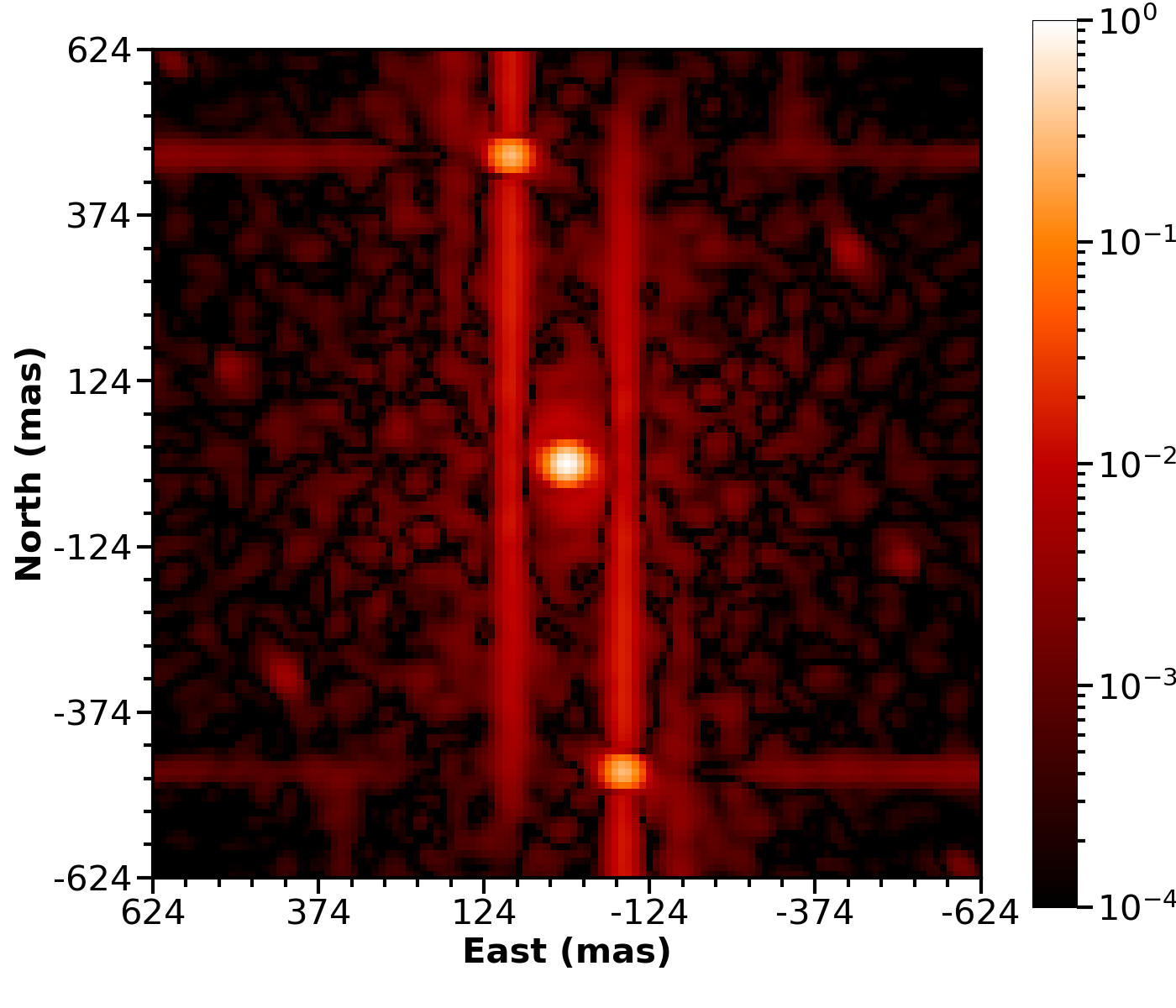}
\includegraphics[width=0.40\textwidth]{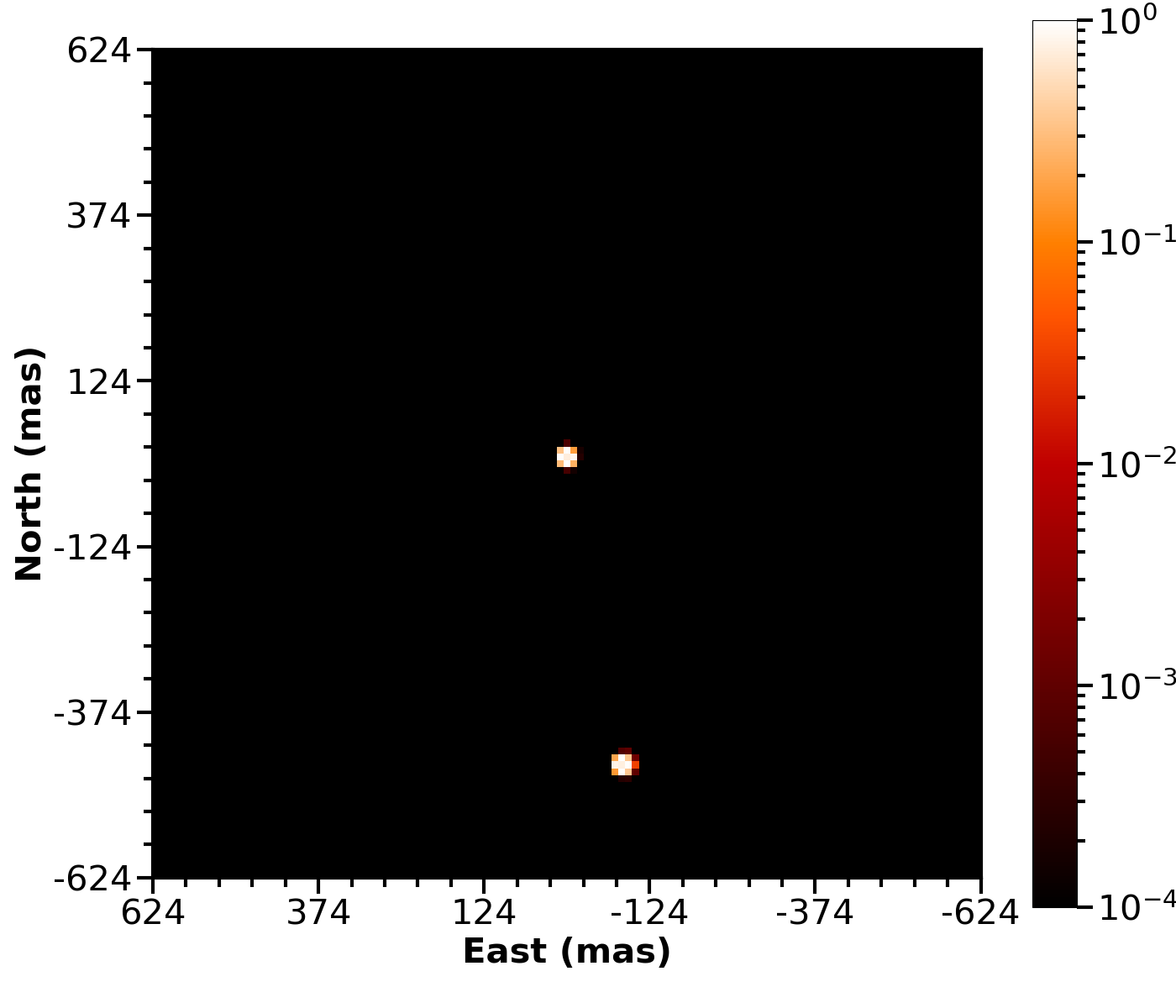}
\includegraphics[width=0.40\textwidth]{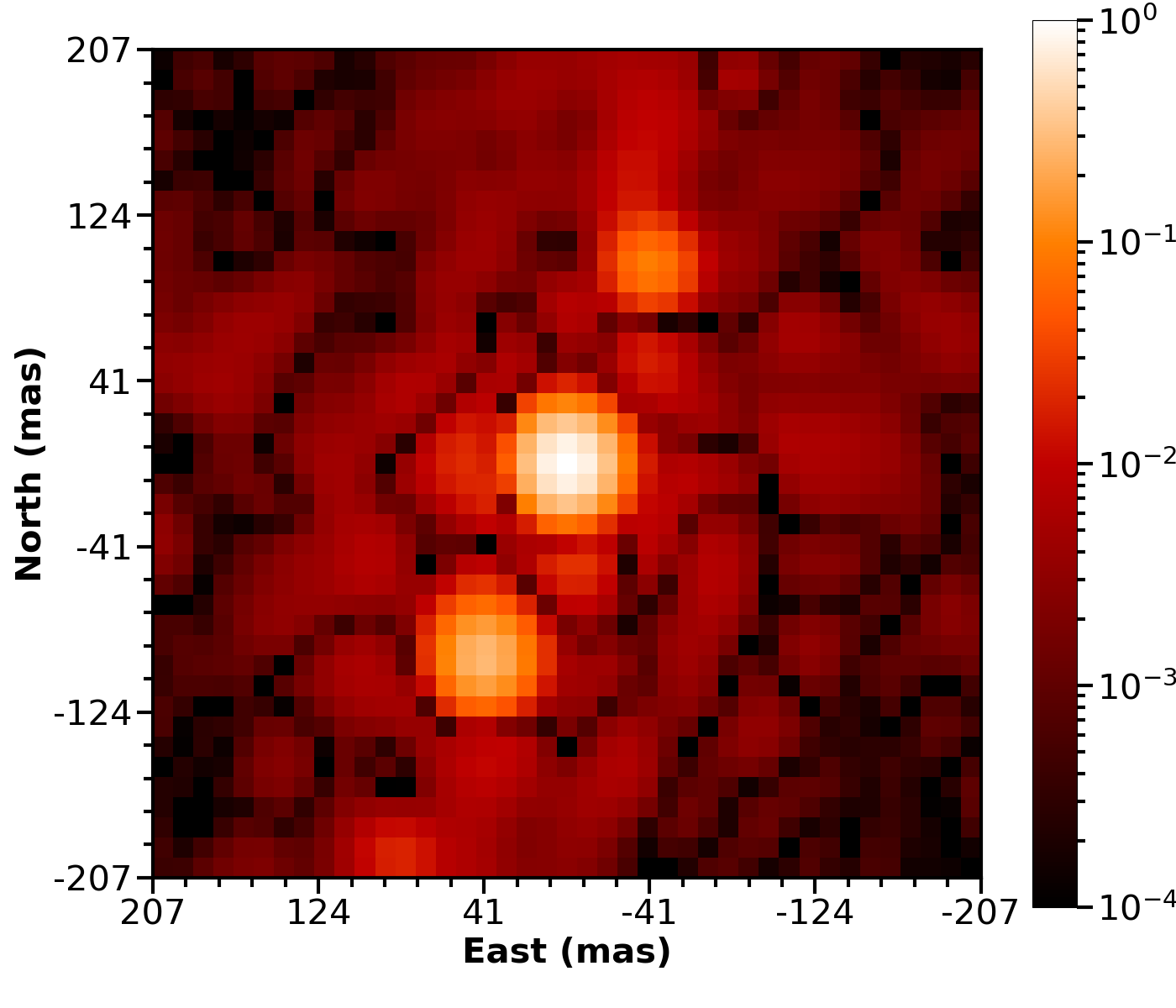}
\includegraphics[width=0.40\textwidth]{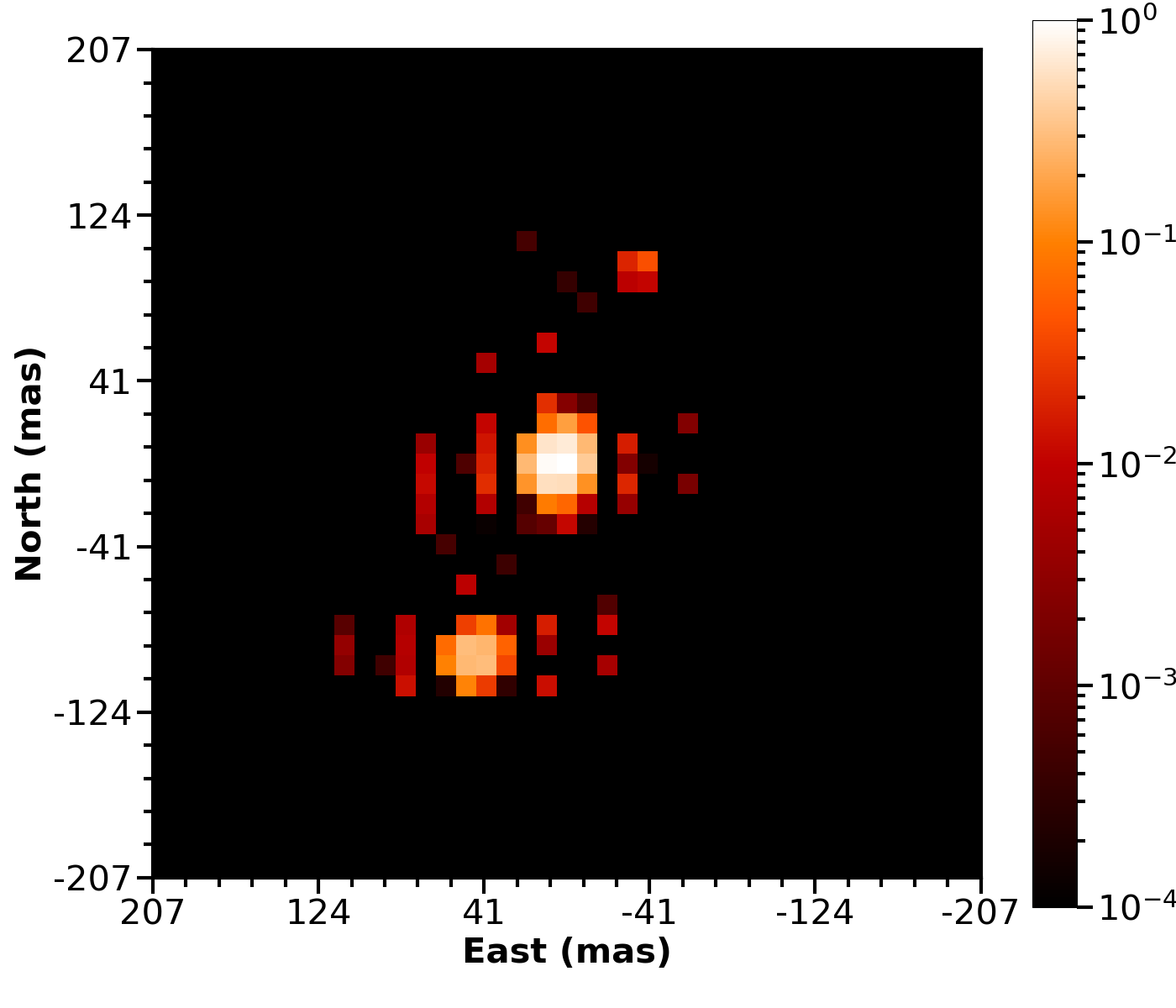}
\caption{
This figure demonstrates the performance of our MFBD algorithm vs.~that of the standard speckle interferometric (SI) reconstructions. Each example reconstructed image uses the same 60 ms speckle data acquired with `Alopeke at the Gemini North 8-m telescope. The left images show the results from our speckle interferometry (SI) Fourier-based analysis of 5000, 60~ms frames for the V=9.91 stellar system TOI-884 (top left), and 3000, 60~ms frames for the stellar systems $\alpha$~Com (middle left) and V819~Her (bottom left). The additional companion seen in V819~Her is not a ghost, see \S3.2.1.
All stars were observed at 832~nm. The right side images show the same data reduced with the MFBD algorithm but using only 1000 for TOI-884 and 500 image frames for $\alpha$~Com and V819~Her.
All images are shown on the same logarithmic intensity scale. Note the higher fidelity and lower background (better contrast - see color bar) of the MFBD reconstructed images. We see in the top two SI reconstructions, the typical 180 degree ghost, a feature that does not occur in the MFBD methods.
The TOI-884 images are shown in a $0.416\times0.416$ arcsecond grid, $\alpha$~Com images are shown in a $1.248\times1.248$ arcsecond grid, and the V819~Her images are shown in a $0.415\times0.415$ arcsecond grid. The TOI-844 and $\alpha$~Com images have a pixel scale of 10.403 mas per pixel whereas the V819~Her images have a pixel scale of 10.393 mas per pixel, and all images are oriented such that North is up and East is to the left.
}
\label{fig:MFBD_results}
\centering
\end{figure}

Due to the nature of Fourier interferometric analysis the reconstruction, the background contrast limits achieved tend to be shallow close to the star, only reaching their full contrast depth near separations of $\sim$0.5$\arcsec$ and beyond \citep[see][and Figure \ref{fig:contrastcurve_all}]{howell:2022}.
It is also axiomatic that speckle interferometric Fourier reconstructions contain a 180 degree ambiguity (a ghost) for detected close companions. However, differentiating the correct companion star from its ghost is often solvable, with good data, using phase information and bispectrum analysis \citep[][see Figure \ref{fig:MFBD_results}]{lohmann:1983}. 

Of interest in this paper are the current highest resolution, deepest contrast speckle instruments available (`Alopeke and Zorro) on the Gemini 8-m telescopes in Hawaii and Chile \citep{scott:2021,howell:2022}. `Alopeke and Zorro are identical dual-channel speckle instruments using EMCCD detectors which have very low noise,  plate scales of $0.01\arcsec$/pixel, and are capable of fast imaging. Speckle imaging using these instruments has been shown to produce an inner working angle at the diffraction limit \citep[$\sim$20 mas;][]{horch:2012,lester:2021} and even some sub-diffraction limit measurements have occurred under conditions of excellent seeing \citep{howell:2021c}. While the necessity of high S/N and good native seeing is crucial for such (sub)diffraction limited inner working angles, modern telescopes such as Gemini are at very good sites and routinely deliver native seeing under 0.5$\arcsec$; a Fried atmospheric coherence length of $r_0 \simeq$~20~cm at 500~nm \citep{fried:1966}.

\section{Multi-frame Blind Deconvolution}

A modern approach to speckle imaging image reconstruction is provided by multi-frame blind deconvolution \citep[MFBD;][]{jefferies:1993,schulz:1993}.The MFBD algorithm is an inverse problem assuming the convolution  of the form $i=O \ast PSF$, where i equals the observed image, O is the true object, and PSF is the point spread function. MFBD estimates both the object and the point spread function (PSF) directly from the observations using computational reconstruction of the wavefront. The use of MFBD provides greater S/N results than typical Fourier speckle reconstruction techniques, but its use for processing speckle data has not yet gained much traction in the astronomy community. The reason for this is mainly due to the high computational cost of the MFBD algorithm. 
A recent development to help speed up the process and yield even better image reconstructions is to provide high-quality initial wavefront estimates to the MFBD algorithm in terms of the PSF shape and x,y location by using a ``compact" version of the MFBD algorithm \citep[i.e., CMFBD;][]{hope:2011,hope:2019,hope:2022}. Figure \ref{fig:MFBD_results} presents a comparison of the reconstructed images produced using the standard SI and the MFBD algorithms.  

\subsection{The Algorithm} \label{sec:MFBD_algorithm}
The MFBD \citep[see e.g.,][]{jefferies:1993,schulz:1993} process proceeds as follows.
For each speckle data frame, we compute the signal-to-noise of each Fourier component ($\frac{|G_k(\mathbf{u})|}{\sigma_N}$); where $|G_k(\mathbf{u})|$ is the Fourier amplitude at spatial frequency {\bf u} and the noise estimate $\sigma_N$ is estimated from the data at spatial frequencies beyond the diffraction-cutoff frequency. At each spatial frequency, the maximum value of the signal-to-noise ratio (S/N) is determined and used to build a map of the maximum Fourier S/N obtained in each frame.
The speckle images are then ranked by their Fourier S/N, where the ``best'' frames are the frames with the highest Fourier S/N.
The advantage of our frame selection algorithm is that it finds the set of data frames that provide the highest S/N at all the sampled Fourier frequencies inside the cut-off frequency. This outcome is not guaranteed using an algorithm based purely on image sharpness such as that proposed by \citet{1986SPIE..627..797M}.

To remove background light, such as the night sky or scattered from a nearby bright star, we take the pixels in the outer three columns and rows in each frame\footnote{A frame is 256X256 pixels.}, find the median value, and subtract this background value from each frame. We then take the best few tens to hundreds of frames, perform a shift-and-add, to yield good approximations of the bright target star x,y position. It is well known that the brightest speckle is not always at the center of the target PSF, thus we use an intensity weighted centroid technique when performing a shift-and-add on speckle images. We use the shift-and-add technique ${\it{not}}$ as a method to find any close companions or provide any accurate representation of the object scene, it simply is an easy way to get starting priors on the approximate location and rough object shape for the bright target. In general, this operation will not reveal any faint companions, only an estimate of the bright star location and extent.

\begin{figure}
\centering
\includegraphics[width=0.30\textwidth]{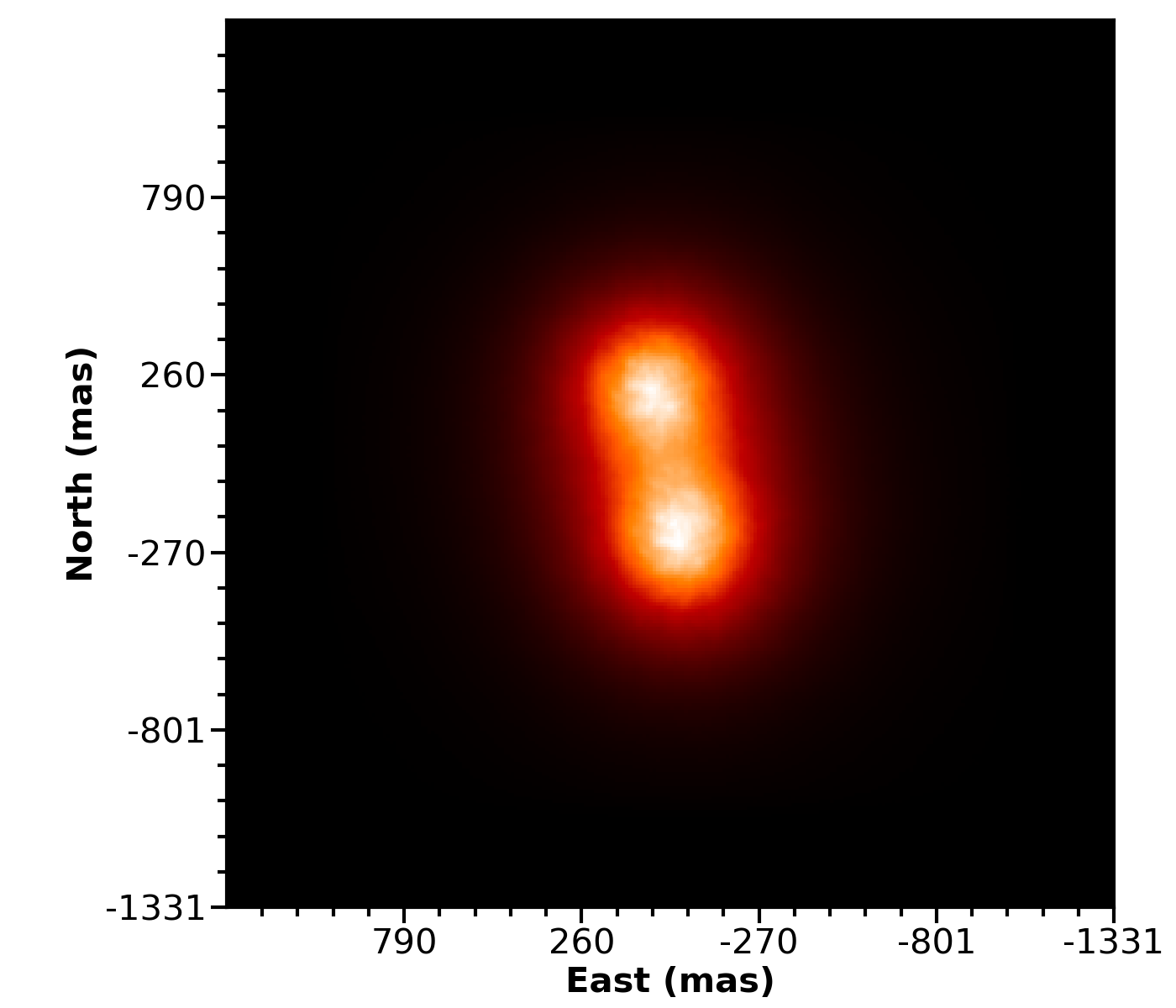}
\includegraphics[width=0.30\textwidth]{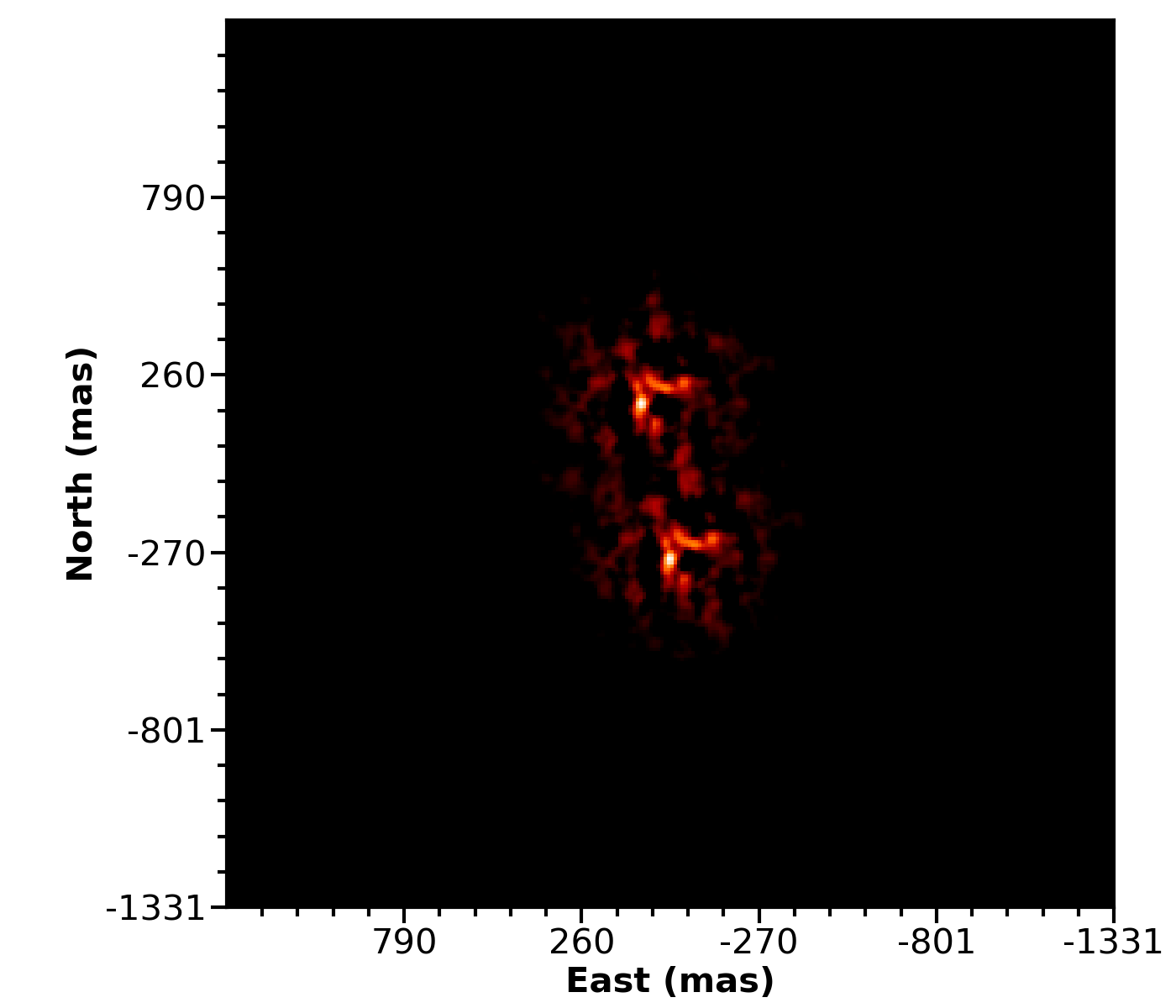}
\includegraphics[width=0.30\textwidth]{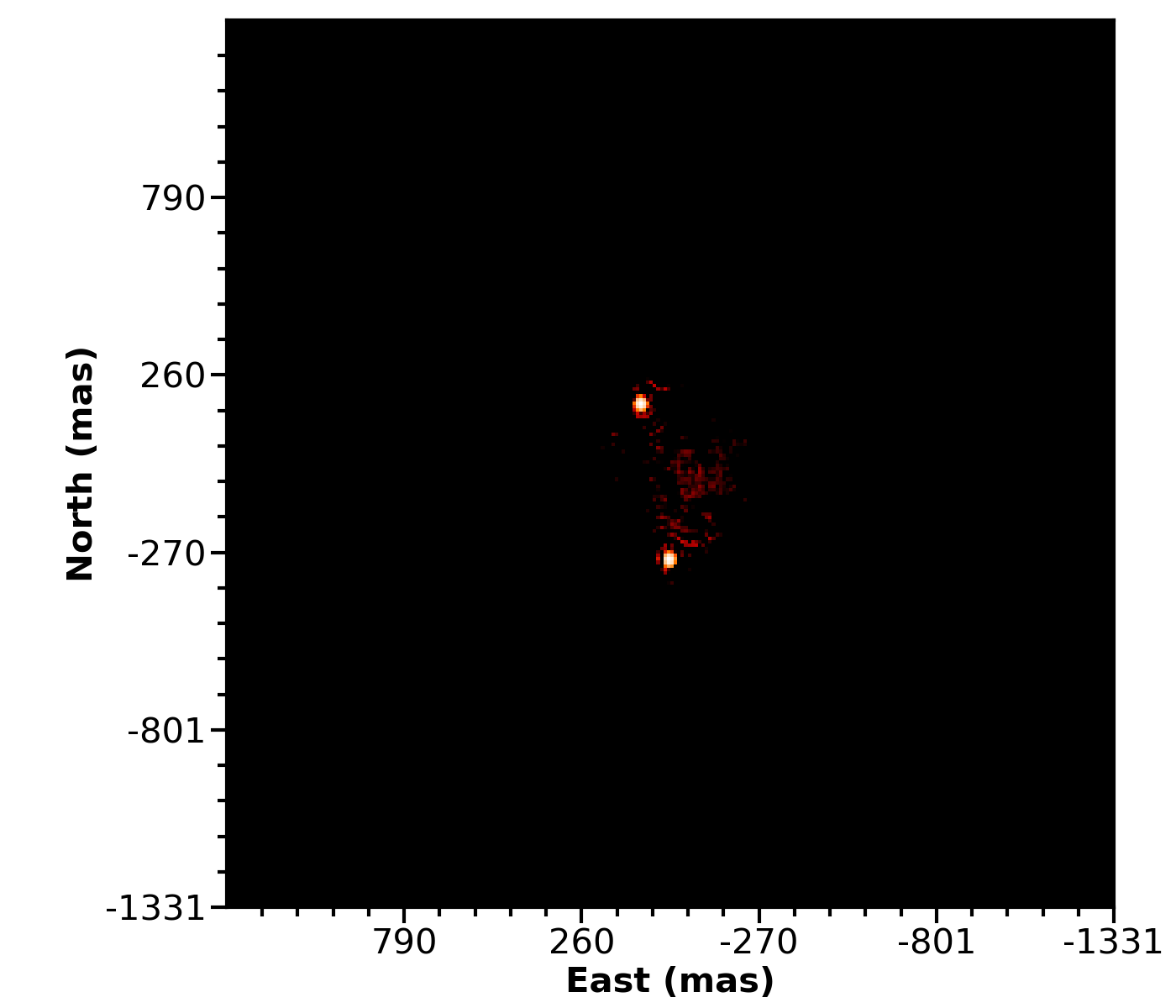}
\includegraphics[width=0.30\textwidth]{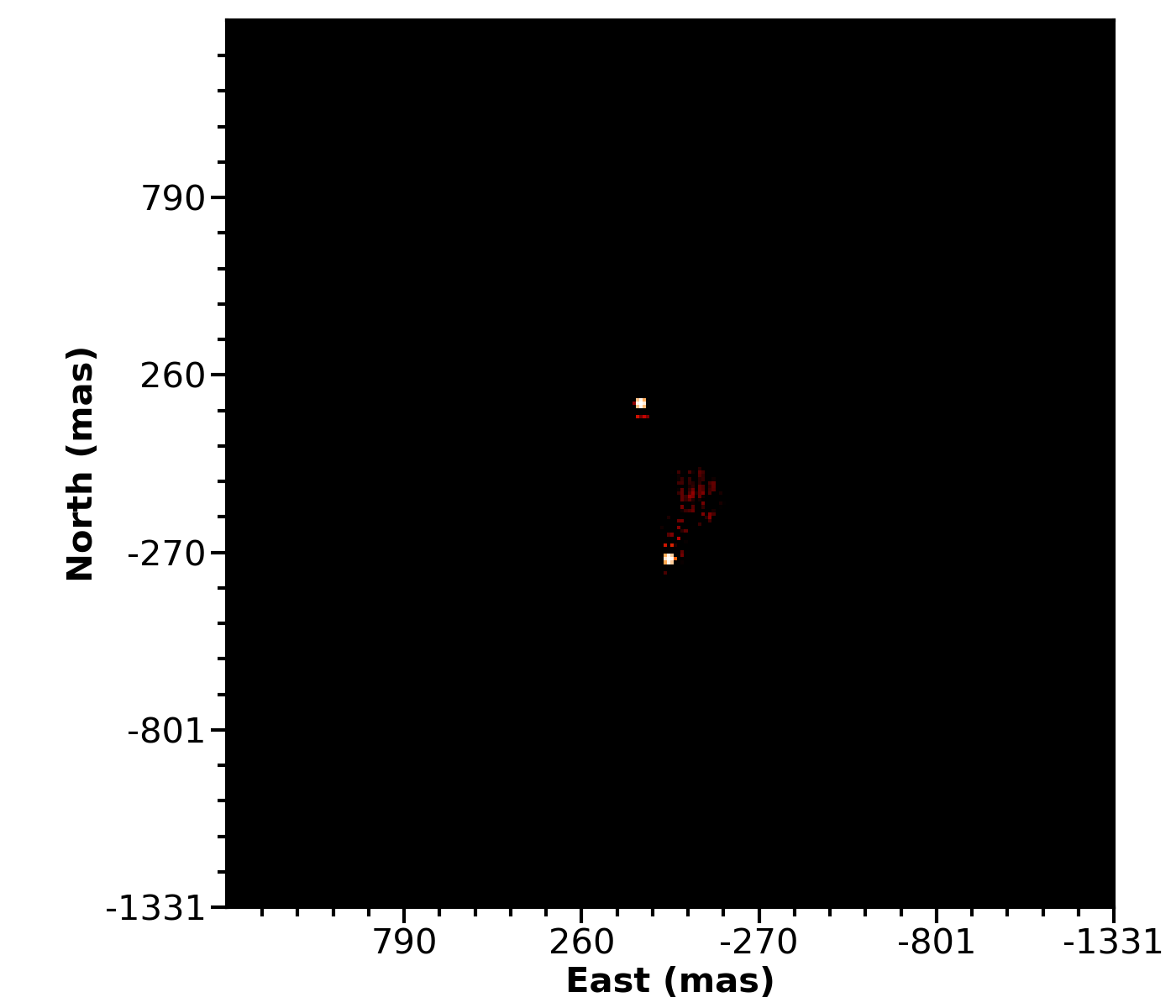}
\caption{
This figure illustrates some steps in the progress throughout the MFBD reconstructions, starting with a preliminary location and PSF shape estimate and ending with the final reconstructed image. The upper left image shows (in linear scale) the starting object image for $\alpha$ Com based on the ``best'' few hundred speckle frames using the shift-and-add method. Note that this example binary, with two nearly equal components, reveals both stars at this step. This would not be the case for a majority of systems, fainter companions would  not be detected at this stage.
The upper middle and upper right images (in logarithmic scale) show the updates of the deconvolved PSF and object scene as the phase of the images is solved for. The lower image (in logarithmic scale) shows the object once complex variables are used allowing the image amplitudes to be incorporated. All images have a pixel scale of 10.403 mas per pixel and North is up and East to the left.
}
\label{fig:kraken_procedure}
\centering
\end{figure}

The reconstruction then follows a three-step process. In the first step, CMFBD is run on a set of the best frames which we call control frames (e.g., first 15 out of all 500 frames; see top left image in Figure \ref{fig:kraken_procedure}) and initialized using the shift-and-add results. Only the pupil tips and tilts are sought, while the amplitudes are kept constant and assumed to be unity. The resulting PSFs are used as initialization for the rest of the frames in the dataset (see top middle image in Figure \ref{fig:kraken_procedure}). In the second step we perform a joint estimation of the object and all the PSFs on all the frames using full MFBD.  The flux ratios for the companion(s) will not be completely accurate at this stage. In this step we only solve for the pupil phase component of the PSF \citep[see top right image in Figure \ref{fig:kraken_procedure};][]{hope:2011}. In the third step, we additionally solve for the pupil amplitudes, further improving the PSF model. The object estimate at the end of this procedure may contain “ring-like” or other low-level blobby structures close to the stars due to static aberrations in the instrument (see bottom image in Figure \ref{fig:kraken_procedure} and Figure \ref{fig:MFBD_results}).

Depending on the total number of frames being used, this full MFBD operation can take many hours up to a few days of computation time (using a modern Intel i9-10700 processor). The computation time can easily be reduced with current technology and in our future implementation, we plan to rewrite various algorithms to increase computational performance, make use of modern day software practices, examine trades such as compute every frame wavefront vs.~average frames together and only compute every N wavefronts and interpolate, make use of higher speed CPUs, and invoke GPU processors. 
Such implementations would provide a factor of 50-100 times in speed-up \citep[e.g., see][]{nelson:2020}, bringing the MFBD analysis time per target into line with that of the current Fourier analysis

In general, any of the ``ring-like'' or blobby structures are not real features, but light from the stellar source that has not fully been integrated into the final PSF, that is, not perfectly modeled. An additional deconvolution can be run using the above PSF model to further refine the modeling and place the light from the erroneous structures back into the stellar PSFs. We note that with MFBD, any static component of the PSF (e.g., due to aberrations in the optical system) will be modeled as part of the object estimate. In the imperfect optical system case, an unresolved source which should reconstruct as a $\delta$-function, will be blurred by a PSF whose morphology depends on the aberrations from the instrument's optics, and time-dependent parameters such as seeing, mechanical changes, and defocus. This additional ``blurring" is clearly seen in Fig.~\ref{fig:MFBD_results}.

\begin{figure}
\centering
\includegraphics[width=0.32\textwidth]{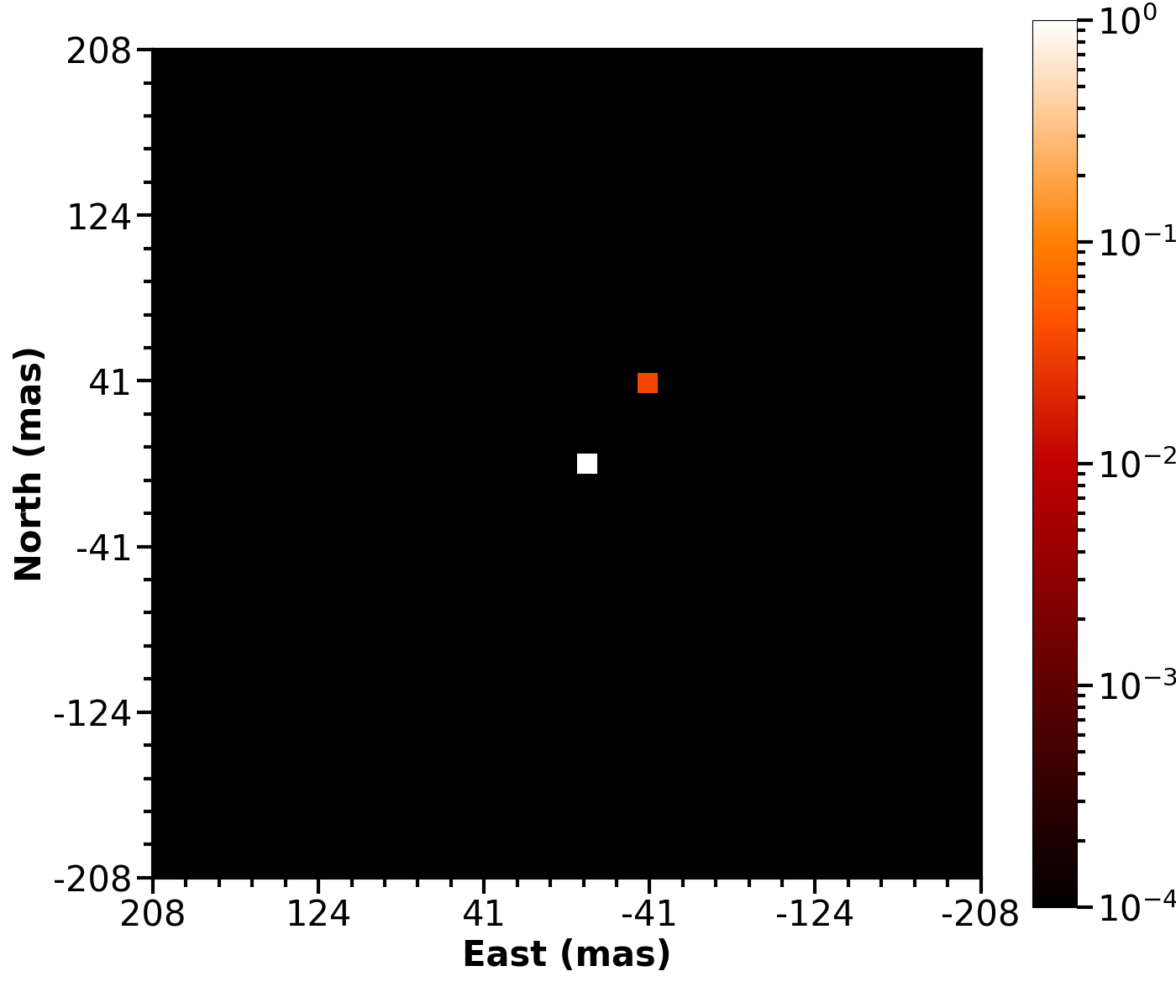}
\includegraphics[width=0.32\textwidth]{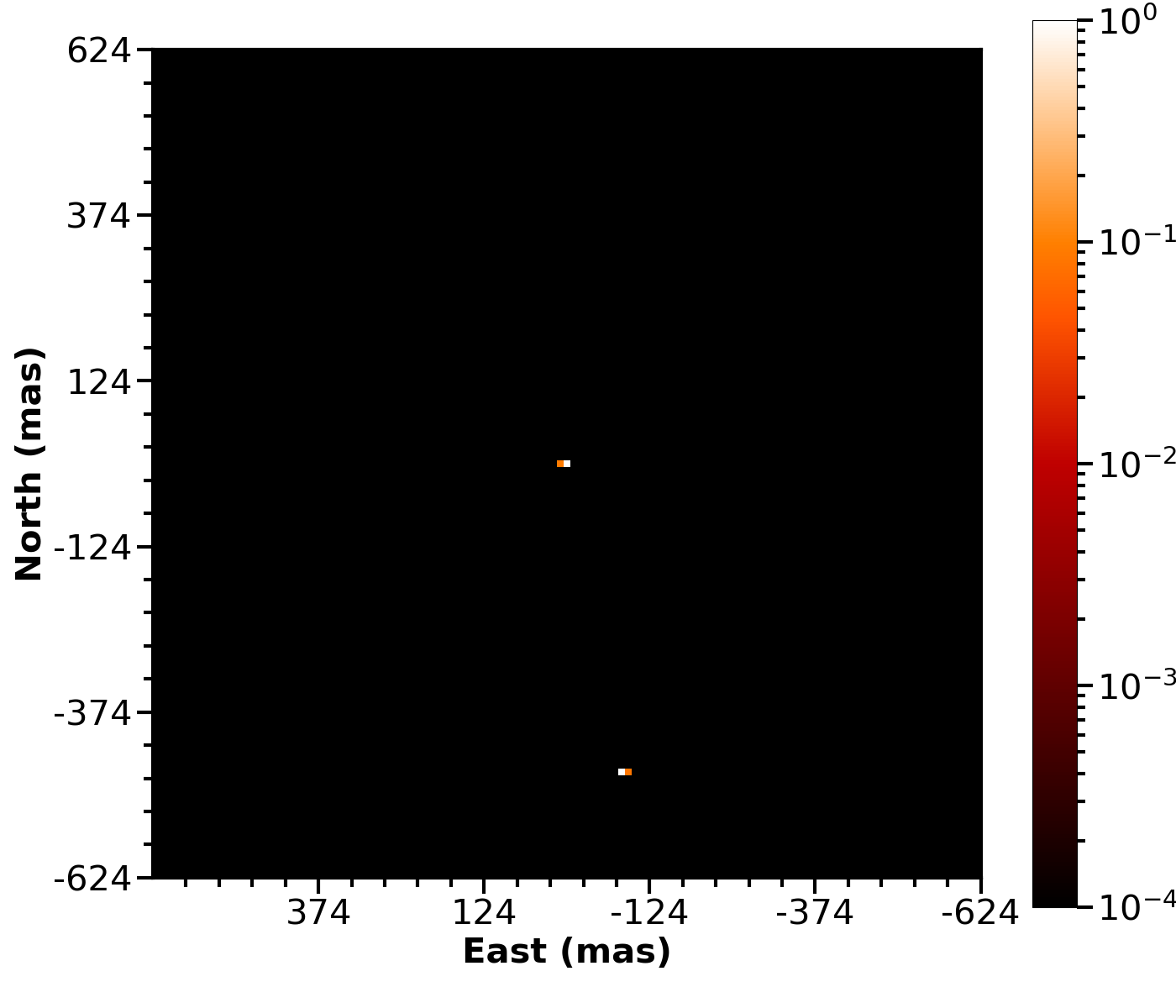}
\includegraphics[width=0.32\textwidth]{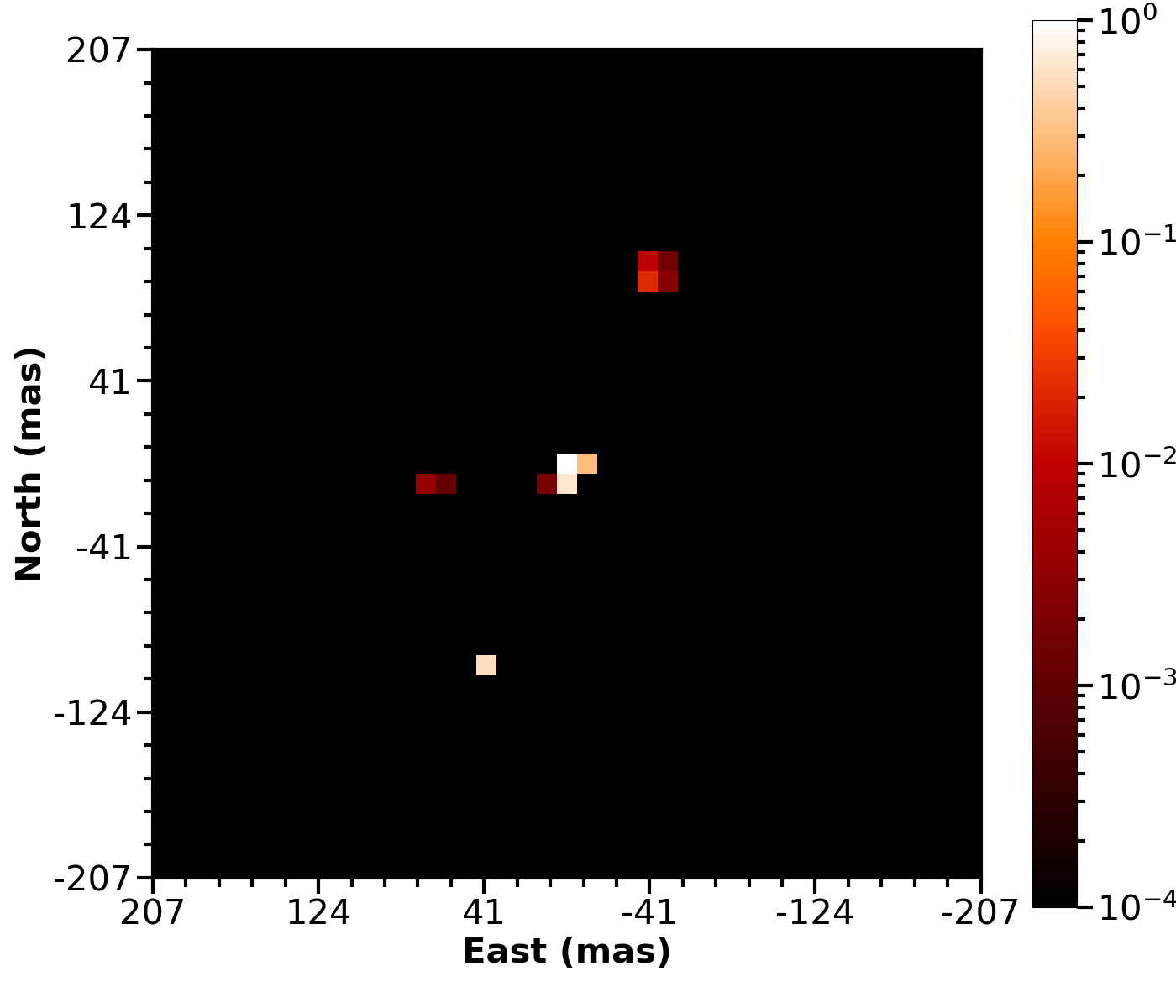}
\caption{
We see here, the final MFBD reconstructed images from the right side of Figure~\ref{fig:MFBD_results} after the removal of the residual blur and aberrations from the instrument PSF. Note that the stars are sub-diffraction limited and will now yield far better astrometric and photometric information as the collected light has been nearly perfectly placed in the scene. All images are shown on the same logarithmic intensity scale. 
The TOI-884 image (left) are shown in a $0.416\times0.416$ arcsecond grid, the $\alpha$ Com (middle) image are shown in a $1.248\times1.248$ arcsecond grid, and the V819~Her image (right) are shown in a $0.415\times0.415$ arcsecond grid. The TOI-844 and $\alpha$~Com reconstructed images have a pixel scale of 10.403 mas per pixel whereas the V819~Her images have a pixel scale of 10.393 mas per pixel, and all images are oriented such that North is up and East is to the left. See \S3.2.
}
\label{fig:SFBD_result}
\centering
\end{figure}

Figure \ref{fig:SFBD_result} shows the object estimates after the removal of an instrumental PSF, including imperfections, for each observation\footnote{Imperfect optical components in `Alopeke have been previously noted, see \cite{scott:2021}.}. Note that the 60 ms images shown in Fig.~\ref{fig:MFBD_results} look ``fuzzy'' as they are the convolution of the pure diffraction limited point source images plus any atmospheric seeing changes, telescope motions (focus, wind, vibration, guiding) as well as the static optical abberations from `Alopeke itself. This ``second deconvolution'', shown in Figure \ref{fig:SFBD_result}, now yields near perfect diffraction limited images.

\subsection{Application of MFBD to Stellar Sources}

\subsubsection{Relative Photometry} \label{sec:MFBD_results}

\cite{horch:2001} and \cite{horch:2011a} discuss the astrophysical uncertainties present in standard Fourier interferometry reconstruction process. These authors show that in good seeing and with high S/N observations, the derived stellar flux ratios are generally good to about $\pm$0.3 mag, becoming larger for very close pairs or widely separated ($>1.0\arcsec$) companions. We perform a comparison to the theoretical flux ratios for three different binary systems between the SI and MFBD techniques.  The theoretical flux ratios are calculated from the individual temperatures and radii of each star, plus the distance of the stellar system, accounting for the `Alopeke 832/40~nm filter transmission.

$\alpha$ Com (HIP 64241) is a bright V~=~4.32 binary system with both stars being F-type main sequence stars \citep[F5V+F6V;][]{tenbrummelaar:2000}. Our SI image reconstruction yields a flux ratio (F$_B$/F$_A$) of 0.71 while our final MFBD reconstructed image provides a flux ratio of 0.99. A theoretical flux ratio calculation for $\alpha$ Com was based on T$_{eff,A}$=6440~K and T$_{eff,B}$=6360~K \citep{eggl:2013}, the radii of R$_{\star,A}$=1.367~R$_{\odot}$ and R$_{\star,B}$~=~1.0281~R$_{\odot}$ derived from the stellar luminosity \citep[L$_{\star,A}$=2.887~L$_{\odot}$ and L$_{\star,B}$=1.553~L$_{\odot}$;][]{eggl:2013} and effective temperature, and a Hipparcos distance of 17.83~pc \citep{vanleeuwen:2007}. With these physical parameters for each star, the flux ratio calculation $\alpha$ Com yields F$_B$/F$_A$ = 0.99 (see middle image in Figure \ref{fig:SFBD_result}). 

V819~Her (HD 157482) is a V~=~5.561 hierarchical triple system with the brighter A component being a G8III star (see \cite{muterspaugh:2006} and \cite{obrien:2011}, and references therein) and the Ba+Bb binary pair as a F2V+G1V \citep{torres:2010} with the system at a distance of 68.8~pc \citep{zasche:2014}. The inner F2V+G1V binary pair is not resolvable through speckle imaging with Gemini's 8.1-m, thus only a single flux ratio calculation can be made. Our SI image reconstruction yields a flux ratio ((F$_{Ba}$+F$_{Bb}$)/F$_A$) of 0.38 while our final MFBD results provides a flux ratio of 0.28.  The theoretical calculations were based on an effective temperature of T$_{eff,A}$=5752~K \citep{bermejo:2013}, R$_{\star,A}$=5.557~R$_{\odot}$ which was calculated from the luminosity \citep[L$_{\star,A}$=30.35~L$\odot$;][]{mcdonald:2012} and effective temperature for the primary star. The eclipsing binary pair physical parameters used for these calculations are the following: R$_{\star,Ba}$=1.87~R$_{\odot}$, R$_{\star,Bb}$~=~1.09~R$_{\odot}$, T$_{eff,Ba}$=6800~K, T$_{eff,Bb}$=5900~K \citep{torres:2010}. The MFBD result for the relative flux difference, 0.28, is close to the theoretical result of 0.25 (see right image in Figure \ref{fig:SFBD_result}). 

We note that the MFBD technique has a much higher sensitivity compared to SI (see Figure \ref{fig:contrastcurve_all}) and is able to detect other components within the image for V819~Her. The first is directly east of the central primary star at a flux ratio of $2.4\times10^{-3}$, very near the sensitivity limit given the number of frames used. Thus, its reality is in question. 
The second component is northwest from the primary star having a flux ratio of $1.8\times10^{-2}$.
This component is coincidentally near 180$^{\circ}$ degrees away in position angle from the Ba and Bb system, but it is many sigma above the detection threshold at 102 mas and thus assumed to be real. Future observations of V819 Her will allow us to obtain deeper imagery and over time, detect (or not) relative motions of the components allowing for an assessment of their reality relationship and relationship to the primary star. 

We also observed the V~=~9.91 planet-hosting system TOI-884 \citep[TIC~167031605, HD~254138;][]{lester:2021}. The MFBD flux ratio is F$_A$/F$_B$ = 0.04 (3.6 magnitudes) while we note that the Fourier reduced archival data gave a flux ratio value is 0.20 (1.7$\pm$0.3 magnitudes).
Given that the companion to TOI-884 was recently discovered, is much fainter than the primary star, and lies at a separation of only $\sim$50 mas, the true flux ratio is unknown. However,
our preliminary findings above for $\alpha$~Com and V819~Her suggest that the flux ratio for TOI-884 from MFBD should be more accurate compared to that from SI (see left image in Figure \ref{fig:SFBD_result}).

While the typical Fourier method relative magnitude mean uncertainty of $\pm$0.1-0.3 magnitudes may not seem important, let us consider an example. If a binary star has a primary G5V star and a companion is detected that is 3$\pm$0.3 magnitudes fainter, that companion can only be assessed as a M0-M3.5 star, a mass difference of 0.6 to 0.25 M$_{\odot}$, or a mass ratio (M$_2$/M$_1$) uncertainty of 0.64 to 0.26. More accurate flux ratio determinations, as those available from MFBD, will be a tremendous help in establishing accurate spectral types and masses for any discovered close companions. A more complete analysis of individual star systems and their properties is not the focus of this paper. \cite{martinez:2023} will provide in-depth comparisons for targets with a wider variety of separations and stellar flux ratios. 

\subsubsection{Achieved Contrast}

All images in Figure \ref{fig:MFBD_results} made use of 60~ms speckle images obtained with `Alopeke and each was reconstructed without any independent phase information (i.e., we have no true wavefront knowledge); yet, the MFBD reconstructions provide more than a factor of ten improvement in terms of S/N and achieved contrast level over the traditional Fourier based SI techniques. To show this more quantitatively, Figure~\ref{fig:contrastcurve_all} presents two different contrast detection curves obtained for $\alpha$ Com in the 832~nm bandpass using each reconstruction technique. The SI contrast curve is obtained by taking the final reconstructed auto-correlated image and measuring concentric rings around the primary star, while ignoring pixels containing the companion \citep{howell:2011}. We also show an average SI curve using data taken with `Alopeke from 2019 to mid-2020 \citep[based on data in][]{scott:2021} in Figure~\ref{fig:contrastcurve_all} as a comparison to other data sets that have varying contrasts.

In order to create the MFBD contrast detection curve, we perform an injection recovery test by adding simulated companions to our speckle frames. We 
start with the modeled $\alpha$~Com primary point source and add various companions with specific delta magnitudes and angular separations. The companion contrast was varied from $10^{-2}$ down to $5\times10^{-4}$ in reference to the primary star and was also modeled as a point source and placed at different angular separations (e.g., 100~mas, 250~mas, 500~mas). We produce a multi-frame series of such models and convolve them with the final model $\alpha$~Com PSFs to generate a sequence of speckle frames to which we add Poisson noise. These steps will allow us to find the sensitivity contrast limit of MFBD using data-driven atmospheric turbulence conditions. 

The comprehensive MFBD image restoration of these simulated speckle images treat them as if they were actual speckle observations. The reduction process is very similar to that described in Section \ref{sec:MFBD_algorithm}, with two minor exceptions. First, instead of starting with the shift-and-add method to yield an initial starting object frame, we assume that the primary star is at the center of the frame (where we placed it) and it is modeled as a delta function. Secondly, 
we skip the CMFBD part of the algorithm and simply run MFBD on all the frames. Contrast level assignment is conducted by measuring the presence of restoration noise (i.e., noise in each difference frame) at each step in the iterative process. The achieved contrast level is established when pixels in the restoration with aberrant flux values reach values comparable to the flux of the secondary in a given frame series. For our set of close companions over the modeled angular separations and flux ratios which had successful discoveries, we fit a contrast level line from the diffraction limit to 1~arcsec as seen in Figure \ref{fig:contrastcurve_all}.

The SI contrast curve shown in Figure~\ref{fig:contrastcurve_all}, a standard reduced data product \citep{howell:2011}, shows the usual slower decline in the near-field approaching 10$^{-3}$ in contrast near 1$\arcsec$ while the MFBD contrast falls precipitously to $\sim10^{-3}$ ($\sim$7.5 magnitudes) at an inner working angle near the diffraction limit, nearly reaching $4\times10^{-4}$ ($\sim$8.5 magnitudes) at 1$\arcsec$. This was achieved with only 500 frames for $\alpha$ Com, a factor of 6 less total frames than typically used in SI. Using the signal-to-noise scaling relation \citep{howell:2022}, we see that MFBD can reach contrasts up to $~\sim8\times10^{-4}$ ($\sim8$ magnitudes) near the diffraction limit and nearly reaching $10^{-4}$ (10 magnitudes) at 1$\arcsec$ (see Figure~\ref{fig:contrastcurve_all}). The MFBD contrast curve is far superior both in terms of near field depth and overall constant performance. Additionally, each companion is more robustly detected and no 180 degree ambiguity ``ghost'' arises from the algorithm. The higher contrast achieved, especially near the diffraction limit, will lead directly to additional and better measurements of close companions and their properties.

\begin{figure}
\centering
\includegraphics[width=0.65\textwidth]{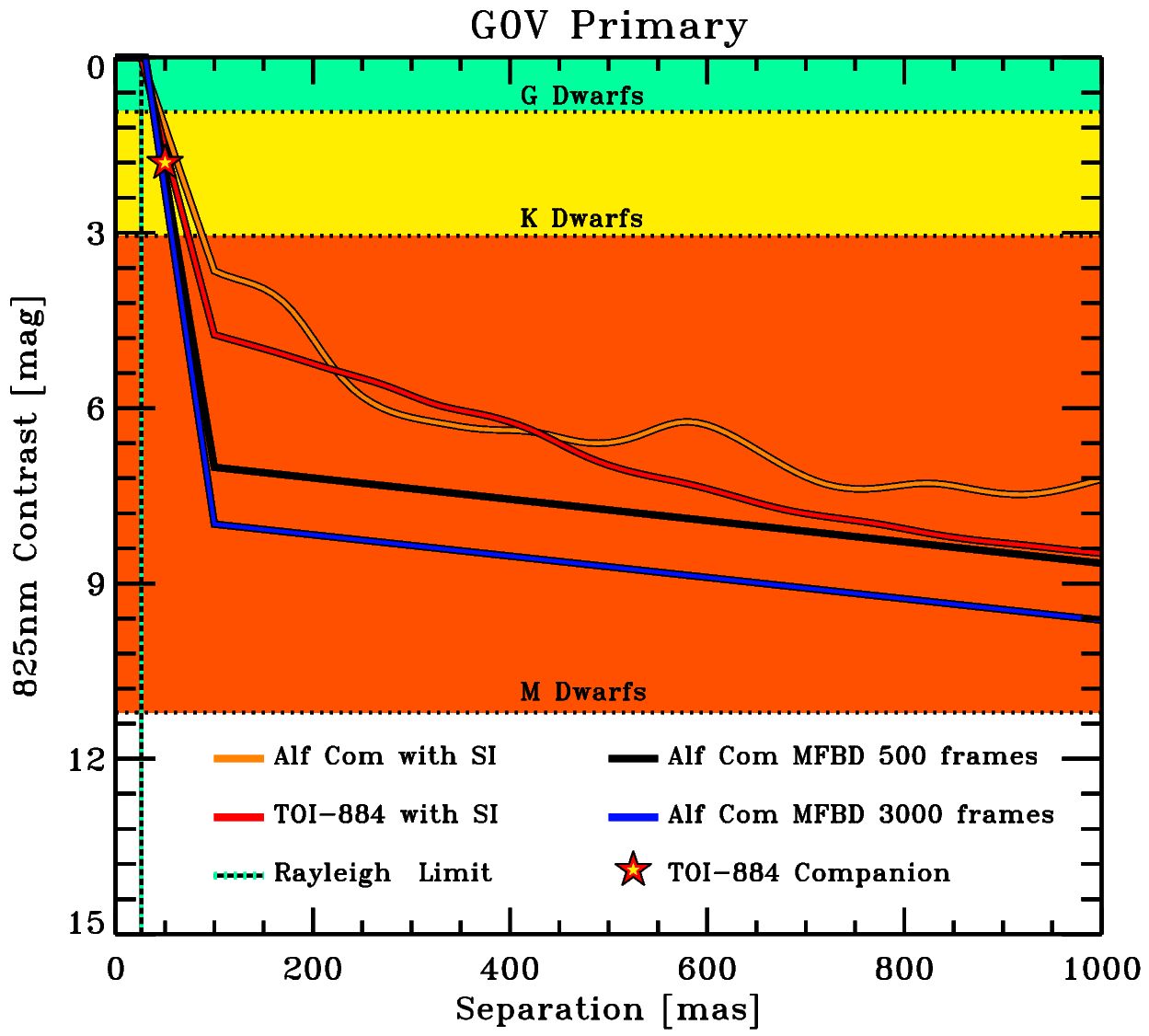}

\caption{The measured 832 nm contrast curves from Fig.~\ref{fig:contrastcurve_all} using the standard Fourier (SI) reconstruction techniques and using the MFBD reconstruction technique.   The vertical dotted green line represents the Rayleigh diffraction limit (1.22$\lambda$/D) at 825~nm for an 8m telescope. The contrast curves are compared to the optical brightness differences between a G0V primary star and the possible range of stellar companions down through the early M-dwarf sequence. Green marks the contrast range of G-dwarfs; yellow marks the contrast range of K-dwarfs; red marks the contrast range of M-dwarfs \citep{PM2013, kirkpatrick2019}. With the new MFBD technique and sufficient total integration time, speckle imaging can reach contrast levels sufficient to detect main sequence companions all the way down through the early-to-mid M-dwarfs on an 8-m telescope. The stellar companion to TOI-884 is marked with the red star and shows that the companion, because of its angular proximity to the primary star ($0.05\arcsec \approx 65au$), was barely detected by the traditional speckle interferometric techniques but is easily detected with the MFBD reduction technique developed here.}.
\label{fig:contrast_starexoplanetplot}
\end{figure}

\section{Relevance to Stellar Astrophysics} \label{sec:companion_detection}

One primary goal of high resolution, high contrast imaging techniques is to detect faint, angularly close companions to a primary target star. Using the new MFBD image reconstruction processing discussed above, the sensitivity and contrast of speckle imaging has been greatly improved. Contrasts of $\approx 10^{-2} - 10^{-3} \approx 5-7$~mag are routinely obtained beyond 0.35 arcsec using our standard Fourier reconstruction techniques with data collected on the 8-m Gemini telescopes. For a typical G0V star, this enables the detection of stellar companions down through the middle of the M-dwarf sequence (see Fig.~\ref{fig:contrast_starexoplanetplot}).  The achieved contrasts are independent of the primary star spectral type (assuming the integration time has been adjusted for the apparent brightness of the target); if the primary star is of a later spectral type (e.g., K-dwarf or M-dwarf), the spectral interferometry contrast can reach further into later spectral types and even into the brown dwarf regime \citep[e.g.,][]{howell:2016}. However, the typical Fourier achieved contrast is strongly dependent upon the separation from the primary star, decreasing to only $\sim 6\times10^{-2} \approx 3$ mag at 0.1\arcsec\ -- nearly 4 times the diffraction limit of the 8m Gemini telescope.

The new data reduction and image reconstruction technique described above achieves contrasts of 5$\times$$10^{-3}$~(8 mag)  starting at essentially the  diffraction limit of the telescope (see Fig.~\ref{fig:contrast_starexoplanetplot}). This improvement in angular resolution and contrast sensitivity over the Fourier techniques enables the detection of nearly 90\% of the entire main sequence below a G0V primary star (Fig.~\ref{fig:contrast_starexoplanetplot}). Using a sufficient number of frames (i.e., long enough total integration time), the sensitivities could reach contrasts of $\sim 10^{-4}$ (9-10 magnitudes), at separations of 0.04\arcsec\ or larger. The limiting inner working angle, shown as vertical line in Figure~\ref{fig:contrast_starexoplanetplot}, represent the Rayleigh diffraction limit (1.22$\lambda$/D) at 825~nm for an 8m telescope. 
  
The traditional Fourier technique reaches a ``flat'' sensitivity level near separations of $0.4 - 0.5\arcsec$ - separations where other measurement techniques, such as Gaia, can also play a role in companion detection. MFBD reaches near-full sensitivity almost immediately - near the wavelength defined full diffraction limit of the telescope. An example of the enhanced improvement is the faint companion detected for TOI-884,  a stellar companion at 0.04\arcsec - barely detected in original SI data reduction and easily detected with the MFBD reduction (Figures \ref{fig:MFBD_results} and \ref{fig:contrast_starexoplanetplot}). The nearly constant sensitivity, starting near the physical diffraction limit of the telescope, enables a substantial increase in the detectable fraction of possible stellar companions.  

\begin{figure}
\centering
\includegraphics[width=0.75\textwidth]{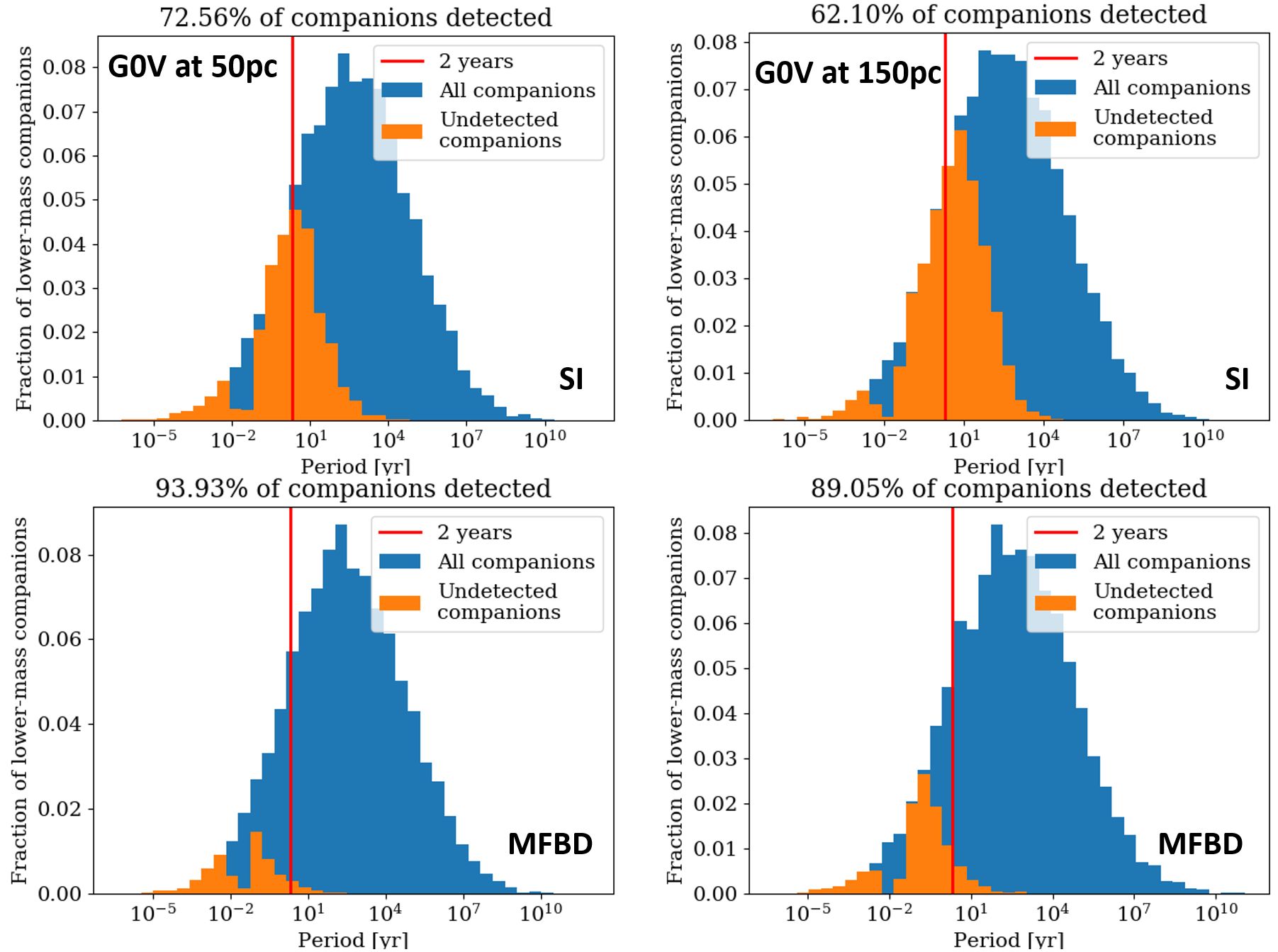}
\caption{The left column is for a generic G0V star at 50pc and the right column is for a generic G0V star at 150pc.  The top row is the expected detection fractions using the traditional Fourier techniques (SI) and the bottom row is results for the new MFBD technique. The blue and orange histograms are the detected and undetected companion distributions, respectively.  The vertical red line marks the 2-year orbital period position for reference. At 50pc, the SI technique detects approximately 70\% of the expected stellar companions, but at 150pc, that fraction drops to 60\%. MFBD enables the detection of $\gtrsim 90\%$ of the expected companions even at 150pc.}
\label{fig:detected_fraction}
\centering
\end{figure}

A Monte Carlo simulation of the total fraction of binary companions expected from a \citet{raghavan2010} companion distribution was performed to investigate the improvement that the new MFBD technique could yield.  The Monte Carlo simulation takes into account the semi-major axis distribution and mass-ratio distribution expected for the local neighborhood population and compares that to the contrast curves from both SI and MFBD \citep[Fig. \ref{fig:contrast_starexoplanetplot}; ][]{lund2020}.  Because of the near-diffraction limited imaging and near-constant sensitivity, MFBD increases the companion detection rate by $30-40\%$ depending on the distance of the star from Earth -- yielding a $\approx 90\%$ detection rate of stellar companions to G-stars (Fig.~\ref{fig:detected_fraction}). The detection fraction is even higher as the primary star become lower in mass (K and M dwarfs).

\section{Summary}

Optical speckle imaging has the advantage of darker skies (than the IR), no need for guide stars (natural or laser), far less expensive and complex instrumentation than IR/AO systems, and is easier setup and use. Speckle observations can be made from 400 nm to 1000 nm using CCD detectors, yielding a spectral energy distribution (SED) for any detected companion across the optical bandpass. Additionally, speckle imaging does not care if the target star is single or multiple, the observations and data reduction processes proceed in the same way. This is not an ability shared by coronagraphic instruments. 

We have shown above that speckle imaging in the optical bandpass using currently available instruments on the Gemini 8-m telescopes, when reduced with our CMFBD/MFBD algorithms, can achieve contrast levels of 10$^{-3}$ to $\sim$10$^{-4}$ from near the diffraction limit out to 1.0 arcsec or more. 
MFBD provides the knowledge of the PSF from the data itself, ending the SI required need for additional observations of point source standards, thus increasing overall observing efficiency. Also the MFBD reconstructed images provide better flux ratios and astrometric information, allowing more precise astrophysical results. We have shown that optical speckle imaging can provide astronomers with a nearly complete census of solar-type main sequence binaries. Such an instrument would provide the highest angular resolution and deepest contrast levels available today in the optical bandpass and would open a new window in astronomical imaging.

\section*{Acknowledgements}
The observations in this paper made use of the High-Resolution Imaging instrument `Alopeke and were obtained under Gemini LLP Proposal Number: GN/S-2021A-LP-105. `Alopeke was funded by the NASA Exoplanet Exploration Program and built at the NASA Ames Research Center by Steve B. Howell, Nic Scott, Elliott P.~Horch, and Emmett Quigley. Alopeke was mounted on the Gemini North 8-m telescope of the international Gemini Observatory, a program of NSF’s OIR Lab, which is managed by the Association of Universities for Research in Astronomy (AURA) under a cooperative agreement with the National Science Foundation. on behalf of the Gemini partnership: the National Science Foundation (United States), National Research Council (Canada), Agencia Nacional de Investigación y Desarrollo (Chile), Ministerio de Ciencia, Tecnología e Innovación (Argentina), Ministério da Ciência, Tecnologia, Inovações e Comunicações (Brazil), and Korea Astronomy and Space Science Institute (Republic of Korea).
This research has made use of the NASA Exoplanet Archive and ExoFOP, which are operated by the California Institute of Technology, under contract with the National Aeronautics and Space Administration under the Exoplanet Exploration Program. Additional information was obtained from the SIMBAD database, operated at CDS, Strasbourg, France. We acknowledge support from AFOSR awards FA9550-14-1-0178 (DAH and SMJ) and FA9550-21-1-0384 (SMJ).

\bigskip

\noindent {\it {Facilities:}} Gemini -  `Alopeke, Zorro

\bigskip
\bigskip
\bigskip

\bibliographystyle{aasjournal.bst}
\bibliography{Exo.bib}







\end{document}